\documentstyle{article}
\begin{document}
\begin{center}
   {\Large \bf Hole-hole contact interaction in the $t$-$J$ model}
\end{center}
\begin{center}
        A. L. Chernyshev\\
  {\em Institute of Semiconductor Physics, 630090 Novosibirsk, Russia}
\end{center}
\begin{center}
        A. V. Dotsenko and O. P. Sushkov\\
  {\em School of Physics, The University of New South Wales\\
   P.O.Box 1, Kensington, NSW 2033, Australia\\
and Budker Institute of Nuclear Physics, 630090 Novosibirsk, Russia}
\end{center}

\begin{abstract}
Using an analytical variational approach we calculate the hole-hole contact
interaction on the N\'{e}el background.  Solution of the Bethe-Salpeter
equation with this interaction gives bound states in $d$- and $p$-waves
with binding energies close to those obtained by numerical methods.  At
$t/J \ge 2-3$ the bound state disappears.  In conclusion we discuss the
relation between short range and long range interactions and analogy with
the problem of pion condensation in nuclear matter.

\end{abstract}

\newpage
\section{Introduction}

Due to connection with high-$T_c$ superconductivity, the problem of mobile
holes in the $t$-$J$ model has received considerable development during
recent years. This model with less than half-filling is defined by the
Hamiltonian
\begin{equation}
\label{1}
  H = H_t + H_J
    = t \sum_{<nm>\sigma} ( d_{n\sigma}^{\dag} d_{m\sigma} + \mbox{H.c.} )
    + J \sum_{<nm>}  \biggl[ S^z_n  S^z_m
        + { \alpha \over 2} ( S^+_n S^-_m + S^-_n S^+_m ) \biggr],
\end{equation}
Where $d_{n\sigma}^{\dag}$ is the creation operator of a hole with spin
$\sigma$ ($\sigma= \uparrow, \downarrow$) at site $n$ of a two-dimensional
square lattice. The operator $d_{n\sigma}^{\dag}$ acts in the Hilbert space
where there is no double electron occupancy. The spin operator is
${\bf S}_n = {1 \over 2} d_{n \alpha}^{\dag}
 ${\boldmath $\sigma$}$_{\alpha \beta} d_{n \beta}$.
$<nm>$ are the neighbor sites on the lattice.
The Hamiltonian (\ref{1}) with $\alpha =1$ corresponds to the $t$-$J$ model,
and for $\alpha= 0$ it is the $t$-$J_z$ model.  Below we set $J=1$.

  At half-filling (one hole per site) the $t$-$J$ model is equivalent to
the Heisenberg antiferromagnet model \cite{Hir5,Gro7} which has long range
N\'{e}el order in the ground state \cite{Oit1,Hus8,Manousakis}.
The problem is the behavior of a system under doping by additional holes.
Strong interaction of mobile holes with localized spins makes this problem
very complicated.

Intense studies has lead to quite firm establishing of
 one particle properties in the $t$-$J$ model.
A single hole is an object with a complex structure
 due to virtual admixture of spin excitations.
It has been shown both analytically \cite{Shr8}-\cite{Sus2}
 and numerically \cite{Tru8}-\cite{Sus1}
 that one hole has a ground state with
 a momentum of ${\bf k}=(\pm \pi/2, \pm \pi/2)$.
The energy is almost degenerate along the line $\cos k_x+\cos k_y=0$
 which is the edge of the magnetic Brillouin zone.
However, even the two hole problem remains controversial
 because there are many sophisticated effects in such systems.

The hole can interact with spin waves. For the N\'{e}el background,
 the spin wave spectrum is gapless according to the Goldstone theorem.
The effective coupling constant for the
 interaction of a hole with long wave length spin excitations for
 $t/J \le 5$ was calculated in our work \cite{Suh3}.
Earlier it had been done in Ref.\cite{Sra9} in the perturbation theory
 limit ($t/J \ll 1$) and for $zS \gg 1$ ($z$ is the number of neighbor
 sites).

If interaction of the Goldstone excitation with particles exceeds
 a certain critical value, the system becomes unstable.
This was understood a long time ago for electron-phonon interaction
 (lattice instability) \cite{Wen1}. Just because of this fact and because
 the hole-magnon coupling constant is large enough, any small but finite
 doping destroys the long range antiferromagnetic order in the $t$-$J$
 model.  It was shown for $t \ll J$ in  Ref.\cite{Shr9}.
For the Hubbard model, the instability
 has been proven in the Hartree-Fock approximation \cite{Sin0}.
For the $t$-$J$ model, the instability of long range antiferromagnetic
 order for  $t/J \le 5$ was demonstrated in our work \cite{Sus3}.
In the same paper, we pointed out the direct analogy of this instability
 with pion condensation in nuclear matter (for review see Ref. \cite{Mig8}).

  In the present work we calculate contact interaction between two
holes on the N\'{e}el background. ``Contact'' means that it is due to
exchange by spin excitations with momentum $q \sim \pi$.
There is no retardation in contact interaction.
We consider this interaction for both
 parallel and opposite directions of the holes spins.
We calculate the vertex function which provides the most general
 description of interaction.
For the most interesting case of opposite spins, there is
 an attraction at $t/J \le 2-3$ which gives a very shallow bound state.
Our result for binding energy agrees with that of Ref. \cite{Ede2}
 obtained by a numerical variational method in a restricted Hilbert space.
It agrees also with recent result of Monte Carlo simulation \cite{Bon2}.
However, there is a disagreement with results of exact diagonalizations
 on finite-size clusters \cite{Bon9},\cite{Has9}-\cite{Rie1}.
Following Refs. \cite{Bon9,Ede2,Pre0}, we suspect that the
 discrepancy is caused by finite-size effects in exact diagonalizations.
Recent calculation \cite{Poilblanc} of the binding energy on larger clusters
 (up to 26 sites) and the scaling with system size indicated a critical value
 of $J/t$ between 0.3 and 0.5, in agreement with our number.

  We base our study on results obtained in \cite{Sus2}.
The suggested trial wave function of a single hole was of the form
$\psi_{{\bf k}\sigma}=h_{{\bf k}\sigma}^{\dag} |0\rangle$,
where $|0\rangle$ is the background
and $h_{{\bf k}\sigma}^{\dag}$ is the creation operator of a dressed hole
\begin{eqnarray}
\label{2}
h_{{\bf k}\uparrow}^{\dag} &=&
  { 1 \over {\sqrt{2N}}} \sum_n (1-\lambda_n)
  \biggl[ \nu_{\bf k} d_{n\uparrow}^{\dag} + S_n^+ \sum_{\delta}
  \mu_{{\bf k},{\delta}} d_{n+{\delta} \downarrow}^{\dag} \biggr]
  \exp (i{\bf k} \cdot {\bf r}_n), \\
h_{{\bf k}\downarrow}^{\dag} &=&
  { 1 \over {\sqrt{2N}}} \sum_n (1+\lambda_n)
  \biggl[ \nu_{\bf k} d_{n\downarrow}^{\dag} + S_n^- \sum_{\delta}
  \mu_{{\bf k},{\delta}} d_{n+{\delta} \uparrow}^{\dag} \biggr]
  \exp(i{\bf k} \cdot {\bf r}_n). \nonumber
\end{eqnarray}
Here $\lambda_n=+1$ for the spin-up sublattice
 and $\lambda_n=-1$ for the spin-down sublattice,
 {\boldmath $\delta$} is a unit vector corresponding
 to one step in the lattice, $N$ is the total number of sites.
The hole energy corresponding to ansatz (\ref{2}) was found
 in Ref.\cite{Sus2} by variational method
\begin{equation}
\label{3}
  \epsilon_{\bf k}=\epsilon_0 + {{\Delta_0}\over{2}}-S_{\bf k}, \hspace{1.cm}
  S_{\bf k} = \biggl [ {\Delta_0^2\over 4} + 4t^2(1+y) - 4t^2(x+y)
  \gamma_{\bf k}^2 \biggr ]^{1/2},  \nonumber
\end{equation} where
\begin{equation} \label{4}
  \gamma_{\bf k} = {1\over 2} (\cos k_x+\cos k_y).
\end{equation}
Expressions for the parameters $\epsilon_0$, $\Delta_0$, $x$, and $y$ in
 terms of ground state correlators are presented in Ref.\cite{Sus2}.
With the normalization of one hole per the lattice of $N$ sites
 the  coefficients in Eq. (\ref{2}) are as follows
\begin{equation}
\label{5}
 \nu_{\bf k} = {1\over 2} \biggl[ {{\Delta_0+2S_{\bf k}}\over{XS_{\bf k}}}
   \biggr]^{1/2},  \hspace{1.cm}
 \mu_{{\bf k},{\delta}}
  = {t\over {[ Y S_{\bf k} (\Delta_0+2S_{\bf k}) ]^{1/2} }}
   [(1+v) - (u+v)\gamma_{\bf k} e^{i{\bf k}{\delta}}].
\end{equation}
The parameters $X$, $Y$, $u$, and $v$ are also expressed in terms of ground
 state correlators \cite{Sus2}.
In the present work we consider two backgrounds:
1) Ising background which is the ground state of the $t$-$J_z$ Hamiltonian
 and has no quantum fluctuations ($S_n^z=\pm {{1}\over{2}}$).
2) N\'{e}el background which is the ground state
 of the Heisenberg model [$\alpha=1$ in the Hamiltonian (\ref{1})].
For these states the numerical values of
 parameters in Eqs. (\ref{3}) and (\ref{5}) are \cite{Sus2}
\begin{eqnarray}
\label{6}
\mbox{I}:\hspace{0.3cm}\epsilon_0=1.0,\ \Delta_0=1.50,\ x=0.00, \ y=0.00, \
          X=1.0,\ Y=1.00, \ u =0.00, \ v=0.00; \nonumber \\
\mbox{N}:\hspace{0.3cm}\epsilon_0=1.2,\ \Delta_0=1.33,\ x=0.56, \ y=0.14, \
          X=0.8,\ Y=0.72, \ u =0.42, \ v=0.12.
\end{eqnarray}

Ansatz (\ref{2}) includes states with a simple hole and states with one
  hopping to a nearest neighbor site ($S^+_n d^{\dag}_{n+\delta \downarrow}$
  can be represented as $-d^{\dag}_{n+\delta \downarrow}
  d_{n\downarrow} d^{\dag}_{n\uparrow}$).
Using the string picture suggested in Ref.\cite{Bul8}, one can say
  that ansatz (\ref{2}) includes only  strings of the minimal length $L=1$.
It is quite natural for $t \ll 1$, but
  it is a rather rough assumption for the Ising background at $t \ge 1$.
However, for the N\'{e}el
  background this ansatz is justified by comparison with numerical simulations.
It turns out that it provides good results up to $t \approx 5$
  (see discussion in Refs. \cite{Sus2,Sus1}).
It means that the hole motion on the N\'{e}el background is mostly
  determined by quantum fluctuations or, in the string language,
  that fluctuations cut strings.
As far as we understand, a similar conclusion was made in Ref.\cite{Sin0}.
Thus we believe that the wave function we use is valid for $t \le 5$.

The plan of the paper is as follows.
In Sec. II we study the interaction of holes with parallel spins.
In Sec. III we consider the case of the holes with opposite spins in
  the Ising limit and the bound state of two holes.
An analysis of the N\'{e}el case is performed in Sec. IV.
Sec. V presents our conclusions.

\section{Interaction of the holes with parallel spins}

We consider a state with two holes
\begin{equation}
\label{7}
 |1,2\rangle
 = h_{{\bf k}_1\uparrow}^{\dag} h_{{\bf k}_2\uparrow}^{\dag} |0 \rangle.
\end{equation}
Using Eq. (\ref{2}), one can easily check that the effective creation
 operators $h^{\dag}_{{\bf k} \sigma}$ obey the usual anticommutation
 relations:
\begin{equation}
\label{8}
  \{ h_{{\bf k}_1\alpha}^{\dag}, h_{{\bf k}_2\beta}^{\dag} \} = 0.
\end{equation}
It means that in the present description the dressed holes are fermions
 and the two hole state (\ref{7}) is well defined.

In the naive perturbation theory,
 the vertex corresponding to scattering $1,2 \rightarrow 3,4$ is
 just the matrix element $\langle 3,4|H|1,2\rangle$.
However in our case, the holes are composite objects and
 we must take into account that the overlapping $\langle 3,4|1,2\rangle$
 is not equal to zero. In this case the vertex is equal to
\begin{equation}
\label{9}
  \Gamma(3,4;1,2)= \langle 3,4| H |1,2 \rangle
    - (\epsilon_1+\epsilon_2) \langle 3,4|1,2 \rangle,
\end{equation}
 where  $\epsilon_i$ is the single particle energy (\ref{3}).
Due to the energy conservation in scattering,
  $\epsilon_1+\epsilon_2 = \epsilon_3+\epsilon_4$.
This procedure with subtraction is well known in the many-body
  perturbation theory (see, e.g., Ref. \cite{Dzu8}).
The physical meaning of Eq. (\ref{9}) is especially  evident
 at $|3,4 \rangle = |1,2 \rangle$ when the interaction is just the
 difference between the total energy and the sum of single-particle energies.

We consider now the Ising background (the $t$-$J_z$ model).
Within the accepted approximation, there is no dispersion for this background
  [see Eqs.(\ref{3}) and (\ref{6})].
Therefore, we may introduce an effective creation operator
  $h^{\dag}_{n \sigma}$ of a hole localized at site $n$.
Equations (\ref{2}-\ref{5}) reduce to
\begin{eqnarray}
\label{10}
&& h^{\dag}_{{\bf k}\sigma}  = \biggl ( {2\over N} \biggr )^{1/2}
  \sum_n h^{\dag}_{n \sigma}  \exp(i {\bf k} \cdot {\bf r}_n), \nonumber\\
&& h_{n\uparrow}^{\dag} = {1\over 2} (1-\lambda_n)
  \biggl( \nu d_{n\uparrow}^{\dag} + \mu S_n^+ \sum_{\delta}
  d_{n+{\delta} \downarrow}^{\dag} \biggr),  \\
&& \epsilon = {7 \over 4} - S ,\hspace{1.0 cm}
  S = \sqrt{9/16+4t^2}, \nonumber\\
&& \nu = {1\over 2} \biggl[ {{3/2+2S}\over S} \biggr]^{1/2}, \hspace{1.cm}
   \mu = {t \over {[S(3/2+2S)]^{1/2}} }.\nonumber
\end{eqnarray}

In the coordinate space the vertex has the form
\begin{equation}
\label{11}
  \Gamma_{\uparrow \uparrow} (m',n';m,n)= \langle m',n'|H|m,n \rangle
    - 2\epsilon \langle m',n'|m,n \rangle,
\end{equation}
where $|m,n\rangle = h_{m\uparrow}^{\dag} h_{n\uparrow}^{\dag} |0 \rangle$
 ($m\not=n$ and $m$ and $n$ are associated with the same sublattice).
One can easily check that matrix elements $\langle m',n'|H|m,n\rangle$ and
 $\langle m',n'|m,n\rangle$ are not equal to zero only for $m'=m$ and $n'=n$.
(There is also a possibility of $m'=n$ and $n'=m$ which,
 due to anticommutation relation (\ref{8}),
 gives the usual fermionic antisymmetrization.)
Obviously, for distant sites $m$ and $n$ the second term
 in (\ref{11}) exactly cancels out the first term.
The interaction is not zero only for the closest possible $m$ and $n$:
 $\Gamma_{\uparrow\uparrow} (n,n+2 \delta_x;n,n+2 \delta_x)$ (Fig. 1a) and
 $\Gamma_{\uparrow\uparrow} (n,n+\delta_x+\delta_y;n,n+\delta_x+\delta_y)$
 (Fig. 1b).
The calculation is straightforward and leads to
\begin{eqnarray}
\label{12}
\Gamma_{\uparrow \uparrow}
  (n,n+2 \delta_x;n,n+2 \delta_x) &\equiv&
  V_{\parallel} = 4t\nu \mu^3(1-4\mu^2)-{1\over 2}\mu^2-4\mu^4+12\mu^6,\\
\Gamma_{\uparrow \uparrow}
  (n,n+\delta_x+\delta_y;n,n+\delta_x+\delta_y) &\equiv&
  V_{\perp} = 16t\nu \mu^3(1-4\mu^2) - \mu^2 - 12\mu^4 + 48\mu^6. \nonumber
\end{eqnarray}
The terms with $t$ are due to the hopping part $H_t$ of the Hamiltonian
 (\ref{1}), and the other terms are due to the $H_J$ part.
In calculations we used that $\nu^2 + 4\mu^2 =1$ and
 $\epsilon=\nu^2+10\mu^2-8t\nu\mu$.

The $\Gamma_{\uparrow \uparrow} (m,n;m,n)$ is in essence the potential
 energy in coordinate space. There is a very weak attraction at $t \le 1$
 which is due to configurations with closed holes minimizing
 $\langle H_J \rangle $ (see, e.g., Fig. 2).
In minima at $t \sim 0.5$, $V_{\parallel} \approx V_{\perp} \approx -0.018$.
Practically within the accuracy of the calculation, it is equal to zero.
At $t \ge 1$ the interaction (\ref{12}) is repulsive.
At $t \gg 1$, $V_{\parallel}\approx {1\over16} (t-{13\over8})$
  and  $V_{\perp} \approx {1\over4} (t-{7\over8})$.

One can easily convert the vertex into momentum representation
 [Eq. (\ref{9}) with $|1,2\rangle = h^{\dag}_{{\bf k}_1 \uparrow}
   h^{\dag}_{{\bf k}_2 \uparrow} |0\rangle$ and
  $|3,4\rangle = h^{\dag}_{{\bf k}_3 \uparrow}
   h^{\dag}_{{\bf k}_4 \uparrow} |0\rangle$]
\begin{equation}
\label{13}
 \Gamma_{\uparrow \uparrow} ({\bf k}_3,{\bf k}_4;{\bf k}_1,{\bf k}_2) =
  {8\over N} \biggl[ V_{\parallel}(\cos^2q_x+\cos^2q_y)+
   V_{\perp}\cos q_x \cos q_y - (3 \rightarrow 4) \biggr] \delta_{12,34}.
\end{equation}
Here ${\bf q} = {\bf k}_1-{\bf k}_3 = {\bf k}_4-{\bf k}_2$;
 $\delta_{12,34} = 1$ if ${\bf k}_1+{\bf k}_2={\bf k}_3+{\bf k}_4$ and
 $\delta_{12,34} = 0$ otherwise.

We expect that the vertex (\ref{13}) calculated for the Ising background
 provides also a reasonable estimate for the case of the N\'{e}el background.
Moreover, we think that for $t \ge 1$,
 due to the arguments presented after Eq. (\ref{6}), it is actually better
 justified for the N\'{e}el background than for the Ising one.
The calculation of the quantum fluctuation correction is rather
 cumbersome and we omit it for the case of parallel spins.
We feel, having an experience with holes of opposite spins,
 that the correction is not large.

\section{Interaction of the holes with opposite spins on the Ising
background. Energy of the bound state}

Similarly to the previous case, it is convenient to use
 coordinate representation (\ref{10}).
The contact vertex is defined by Eq. (\ref{11}) with the state
 $|m,n\rangle =h_{m\uparrow}^{\dag} h_{n\downarrow}^{\dag} |0\rangle$.
For distant sites $m$ and $n$, the second term in
 (\ref{11}) exactly cancels out the first term and the interaction vanishes.
As in the case of parallel spins, there is a potential-like
 term $\Gamma_{\uparrow \downarrow}(n,n+\delta_x;n,n+\delta_x)$ (Fig. 3a).
In addition, there appear effective hopping of clusters
 $\Gamma_{\uparrow \downarrow} (n,n+\delta_y;n,n+\delta_x)$ (Fig. 3b) and
 $\Gamma_{\uparrow \downarrow} (n,n-\delta_x;n,n+\delta_x)$ (Fig. 3c)
 which turn out to be equal.
Simple calculation gives
\begin{eqnarray}
\label{14}
&\Gamma_{\uparrow \downarrow} (n+\delta_x;n,n+\delta_x) \equiv
  U = 4t\nu \mu (1-5\mu^2+4\mu^4)
    - {1\over 4}-4\mu^2 + {27\over 2}\mu^4 - 12\mu^6,    \nonumber\\
&\Gamma_{\uparrow \downarrow} (n,n-\delta_x;n,n+\delta_x) =
 \Gamma_{\uparrow \downarrow} (n,n+\delta_y;n,n+\delta_x) \equiv T, \\
&T = 2t\nu \mu (1-9\mu^2+32\mu^4) - \mu^2(1-4\mu^2)(1-12\mu^2).\nonumber
\end{eqnarray}
Again, the terms with $t$ are due to the hopping part of the Hamiltonian
 (\ref{1}), and the other terms are due to the $H_J$ part.
To be accurate, we should say that within the wave function
 (\ref{10}) in the $t$-$J_z$ model there are more distant configurations
 which contribute to hole-hole interaction (Fig. 4).
However, the corresponding values of interaction potential
\begin{equation}
\label{15}
 \Gamma_{\uparrow \downarrow} (n,n+2\delta_x+\delta_y;n,n+2\delta_x+\delta_y)
  = 3\Gamma_{\uparrow \downarrow} (n,n+3\delta_x;n,n+3\delta_x)
  = -{3\over 4}\mu^4
\end{equation}
are very small and we neglect them.

Now we can consider the bound state of two holes.
It is quite obvious that in the present approximation the wave function
 of such a state with the total momentum $\bf p$ must be of the form
\begin{equation}
\label{16}
\psi_{\bf p} = \sqrt{2\over N} \sum_{n,m}
  \chi ({\bf r}_n-{\bf r}_m) h_{m\uparrow}^{\dag} h_{n\downarrow}^{\dag}
  |0\rangle \exp [ i {1\over2} {\bf p} \cdot ({\bf r}_n + {\bf r}_m)].
\end{equation}
The equation for bound state is given by variation of
$\langle \psi_{\bf p}|(H-E)|\psi_{\bf p} \rangle$ with respect to the
coefficients $\chi ({\bf i})$. If we separate the binding energy $\Delta$
($E=2\epsilon+\Delta$), the equation becomes of the form
\begin{equation}
\label{17}
 \delta \biggl( \langle \psi_{\bf p}| (H-2\epsilon) |\psi_{\bf p}\rangle
   - \Delta \langle \psi_{\bf p}|\psi_{\bf p}\rangle  \biggr) = 0.
\end{equation}
The matrix elements $\langle m',n'|(H-2\epsilon)|m,n\rangle$ are given by
 Eqs. (\ref{14}).
In the term $\Delta \langle \psi_{\bf p}|\psi_{\bf p}\rangle$ we set
 $\langle m',n'|m,n\rangle=\delta_{mm'}\delta_{nn'}$.
Actually, the correction to the last equality is small,
 and therefore it gives only a small correction to $\Delta$.
To avoid misunderstanding, we should note that in the effective
 interaction (\ref{12},\ref{14}) we calculated the overlapping
 $\langle m',n'|m,n\rangle$ exactly because $2\epsilon$ is much larger than
 the binding energy $\Delta$. Further calculation is straightforward.
Only the coefficients $\chi$({\boldmath $\delta$}) corresponding
 to the closest positions of clusters are not equal to zero.
The obtained coefficients correspond to $p$- and $d$-wave  states
 which turn out to be degenerate.
The energy is minimal at ${\bf p}=0$, and the binding energy is equal to
\begin{equation}
\label{18}
 \Delta = U-2T = 16t\nu \mu^3(1-7\mu^2) -{1\over 4}-2\mu^2-18.5\mu^4+84\mu^6.
\end{equation}

  Now we would like to derive Eq. (\ref{18}) using the momentum representation.
For the $t$-$J_z$ model, it is just another mathematical way.
Nevertheless, it is useful because for the $t$-$J$ model, which we are
 mainly interested in, the coordinate representation is not convenient.
Taking $|1,2 \rangle = h_{{\bf k}_1\uparrow}^{\dag}
 h_{{\bf k}_2\downarrow}^{\dag} |0 \rangle$ and
 $|3,4 \rangle = h_{{\bf k}_3\uparrow}^{\dag}
 h_{{\bf k}_4\downarrow}^{\dag} |0 \rangle$ in Eq.(\ref{9}) and
 using Eqs. (\ref{10}), one can easily transform the interaction
 (\ref{14}) into momentum representation [cf. with Eq. (\ref{13})]
\begin{eqnarray}
\label{19}
&&\Gamma_{\uparrow \downarrow} ({\bf k}_3,{\bf k}_4;{\bf k}_1,{\bf k}_2)
  = {8\over N} \biggl[ A\gamma_{\bf q}
  + B (\gamma_1 \gamma_3 + \gamma_2 \gamma_4) \biggr] \delta_{12,34},\\
&&A = U-2T = 16t\nu \mu^3(1-7\mu^2) - {1\over4}-2\mu^2-18.5\mu^4+84\mu^6,
  \nonumber\\
&&B = 4T = 8t\nu \mu(1-9\mu^2+32\mu^4) - 4\mu^2(1-4\mu^2)(1-12\mu^2).\nonumber
\end{eqnarray}
Here $\gamma_i \equiv \gamma_{{\bf k}_i}$.
The simple structure of the vertex (\ref{19}) reflects the simplicity of
 quasiparticle ansatz (\ref{10}).
The quasiparticles interact only when they are at the nearest neighbor sites
 which gives only $\gamma_{\bf q}$ and $\gamma_i \gamma_j$ terms.

Let $g_{\bf k}$ be the wave function of a pair
 with the total momentum ${\bf p}=0$:
\begin{equation}
\label{20}
\psi ({\bf r}_n,{\bf r}_m)={{1}\over{N^{3/2}}} \sum_{\bf k} g_{\bf k}
 h^{\dag}_{{\bf k}\uparrow} h^{\dag}_{- {\bf k} \downarrow} |0\rangle
  \exp[i {\bf k} \cdot ( {\bf r}_n - {\bf r}_m )].
\end{equation}
This function obeys the Bethe-Salpeter (BS) equation
\begin{equation}
\label{21}
 (E-2\epsilon_{\bf k}) g_{\bf k} = \sum_{\bf k'}
   \Gamma_{\uparrow \downarrow}
   ({\bf k},-{\bf k};{\bf k'},-{\bf k'}) g_{\bf k'},
\end{equation}
where the summation is carried out over the Brillouin zone
 ($\cos k'_x+\cos k'_y \ge 0$).
For the Ising case $\epsilon_{\bf k} = \epsilon$ [see Eq. (\ref{10})].
Using the vertex (\ref{19}), one can easily find that  there are two
 degenerate solutions of Eq. (\ref{21}) corresponding to two symmetries
\begin{eqnarray}
\label{22}
 d\mbox{-wave}: \ g_{\bf k} &=& \sqrt{2} (\cos k_x-\cos k_y),\\
 p\mbox{-wave}: \ g_{\bf k} &=& 2 \sin k_x. \nonumber
\end{eqnarray}
The $B\gamma_i \gamma_j$ terms cancel in the process of integration.
The value of binding energy is dertermined by the $\gamma_{\bf q}$ part
$\Delta=A=U-2T$ which exactly agrees with Eq. (\ref{18}).

Plot of the hole-hole binding energy $\Delta$ in the $t$-$J_z$ model as a
 function of $t$ is presented in Fig. 5.
The result at $t=0$ ($\Delta=-1/4$) can be very easily understood.
Two simple holes bound together have one less antiferromagnetic
 bond broken than when they are apart.
As $t$ increases, holes tend to part to gain delocalization energy.
For $t \le 2$ there is a reasonable agreement between our result and
 that of Ref.\cite{Ede2}. However for larger $t$, our curve lies
 substantially lower than in Ref.\cite{Ede2}.
Keeping in mind that our wave function on the Ising background (\ref{10})
 is not very good for $t \ge 1$ [see discussion after Eq. (\ref{6})],
 we would not like to insist on our result.
Let us discuss the point in more detail.
In our calculations for $t-J_z$ model we make in essence an expansion in
${t\over zJ}$, where $z=4$ is the number of neighbor sites.
Therefore, one can expect it to be valid up to $t/J \sim 4$.
Unfortunately at large $t$, there takes place a strong
 compensation in the hopping part of the binding energy (\ref{18})
 $\Delta = U-2T \sim {t\over 2}(1-7\mu^2) \rightarrow {t\over 16}$.
It means that this contribution is asymptotically $1/z$.
Corrections from including longer strings
into the wave function (\ref{10}) are $1/z$ as well. Therefore they
are very important for this value. The situation is different in the
$t$-$J$ model and now we come to this problem.

\section{Interaction of the holes with opposite spins on the N\'{e}el
background. Energy of the bound state}

Since we are finally interested in calculation of the binding energy,
 let us look at the BS equation (\ref{21}).
The N\'{e}el case differs from the Ising one in two ways:
1) There is a not-trivial dispersion $\epsilon_{\bf k}$ [Eq.(\ref{3})].
2) The vertex function $\Gamma$ is renormalized by quantum fluctuations.
Generally speaking, the calculation of fluctuation correction to $\Gamma$
 is straightforward:
One should only substitute the wave functions (\ref{2}) into Eq. (\ref{9}).
As we have pointed out in discussion after Eq. (\ref{6}),
 the wave function (\ref{2}) on the N\'{e}el background is actually more
reliable than on the Ising one. Unfortunately straightforward calculation
of $\Gamma$ is extremely cumbersome because there appear many complicated
correlators. Thus we have to make some simplifications.

1) For the Ising case the wave function of bound state (\ref{22}) is
concentrated near the edge of the Brillouin zone ($\gamma_{\bf k}=0$).
We will prove that for the N\'{e}el background this concentration
is even stronger. (The same conclusion can be made from the plots of wave
functions presented in Ref. \cite{Ede2}.) The reason of this effect is
quite evident: The bottom of the single-hole band
(\ref{3}) is exactly at the edge of the Brillouin zone.
Using this fact, we will consider the vertex
$\Gamma_{\uparrow \downarrow} ({\bf k}_3,{\bf k}_4;{\bf k}_1,{\bf k}_2)$
only for momenta lying at the edge of the Brillouin zone ($\gamma_i=0$).

2) Another simplification is connected with calculation of background
spin correlators \\
$\langle 0| S^{+}_{n} S^{-}_{m} S^{z}_{l} \dots|0\rangle$.
We make an expansion in the parameter $\alpha$ of the Hamiltonian (\ref{1})
 and restrict ourselves to the first order.
Although the physical value is $\alpha=1$, this can be quite a reasonable way
 to estimate the influence of quantum fluctuations.
For example, for the transverse nearest neighbor correlator this gives
 $\langle 0| S^{+}_{n} S^{-}_{n+\delta} |0\rangle \equiv 2q_{1}
  = -\frac{\alpha}{6} \rightarrow -0.167$ which is in good
 agreement with exact numerical value. In the first order in $\alpha$, it is
 the only correlator which differs from its Ising value.

3) The last simplification is the following.
We will calculate quantum fluctuation correction only in the $H_t$ part of the
vertex.  Actually, in the Ising limit there is no compensation in
$H_J$ contribution into the vertex
(\ref{19}) and into the binding energy (\ref{18}).
Therefore, one should to expect only small correction from fluctuations.
It is not the case for the $H_t$ contribution where happens a strong
compensation: $\Delta= U-2T \sim {t\over 2} (1-7\mu^2) \rightarrow {t\over 16}$
(see discussion in the end of the previous section).

  Calculation with the given above simplifications is still cumbersome but
straightforward. Using Eqs. (\ref{2}) and (\ref{9}), we get the correction
\begin{equation}
\label{23}
\Delta \Gamma_{\uparrow \downarrow} ({\bf k}_3,{\bf k}_4;{\bf k}_1,{\bf k}_2)
  = -2 q_1 t {8\over N} \biggl[ \nu \mu^3
  (14\gamma_{\bf q}+\gamma_{{\bf k}_1+{\bf k}_3}+\gamma_{{\bf k}_2+{\bf k}_4})
  +(16\nu^3 \mu^3 - 56\nu \mu^5)\gamma_{\bf q}  \biggr]  \delta_{12,34},
\end{equation}
where $2q_1 = \langle 0| S^{+}_{n} S^{-}_{n+{\delta}}|0 \rangle
= -\frac{\alpha}{6}$ is the nearest neighbor correlator.
Clearly, the correction (\ref{23}) is repulsive.
The coefficients $\nu$ and $\mu$ in Eq. (\ref{23})
correspond to the Ising limit (\ref{10}) since the expansion in $\alpha$ is
carried out explicitly.
The $\nu \mu^3$ term in Eq. (\ref{23}) comes from
 expansion of $\langle 3,4|H_t|1,2\rangle$ in Eq. (\ref{9}),
 the $\nu^3 \mu^3$ term comes from overlapping $\langle 3,4|1,2\rangle$, and
 the $\nu \mu^5$ term comes from expansion of energy $(\epsilon_1+\epsilon_2)$.
The $(\gamma_{{\bf k}_1+{\bf k}_3} + \gamma_{{\bf k}_2+{\bf k}_4})$ part
 arises from interaction of the holes at next-nearest neighbor sites.
Its relatively small value reflects the importance of
 quasiparticles interaction at the nearest neighbor sites.

  Adding the quantum fluctuation correction (\ref{23}) to the Ising value
of vertex (\ref{19}), we get contact hole-hole vertex for the $t$-$J$ model
\begin{eqnarray}
\label{24}
&&\Gamma_{\uparrow \downarrow}' ({\bf k}_3,{\bf k}_4;{\bf k}_1,{\bf k}_2)
 = {8\over N}\biggl[ A'\gamma_{\bf q}
 + B(\gamma_1 \gamma_3 + \gamma_2 \gamma_4)
 + {C\over 2} (\gamma_{{\bf k}_1+{\bf k}_3}+\gamma_{{\bf k}_2+{\bf k}_4})
    \biggr] \delta_{12,34},\\
&&A' = A + 10\alpha t \nu^3 \mu^3, \hspace{1cm}
  C = {2\over 3}\alpha t \nu \mu^3.\nonumber
\end{eqnarray}
Expressions for $A$ and $B$ are presented in Eq. (\ref{19}). In transformation
from Eq. (\ref{23}) to Eq. (\ref{24}) we take into account the normalization
condition $\nu^2+4\mu^2=1$.

  Let us consider two hole bound state with zero momentum ${\bf p}=0$.
If we expand the single hole dispersion (\ref{3}) near the band bottom
\begin{equation}
\label{25}
 \epsilon_{\bf k} = \epsilon_{bot} + {1\over 2}\lambda \gamma_{\bf k}^2,
 \hspace{1 cm}
 \lambda = {{4t^2(x+y)} \over {[\Delta_0^2/4 + 4t^2(1+y) ]^{1/2} } },
\end{equation}
the BS equation (\ref{21}) can be rewritten in the form
\begin{equation}
\label{26}
  (\Delta-\lambda \gamma_{\bf k}^2) g_{\bf k} = \sum_{\bf k'}
  \Gamma_{\uparrow \downarrow}'
   ({\bf k},-{\bf k};{\bf k'},-{\bf k'}) g_{\bf k'},
\end{equation}
where $\Gamma_{\uparrow \downarrow}'$ is given by Eq. (\ref{24}).
Let us first treat $\lambda$-term perturbatively.
At $\lambda=0$, in spite of appearance of
$(\gamma_{{\bf k}_1+{\bf k}_3}+\gamma_{{\bf k}_2+{\bf k}_4})$ term in the
vertex,
the solutions (\ref{22}) still satisfy Eq. (\ref{26}) with binding energies
\begin{equation}
\label{27}
 \Delta_p^{(0)} = a_p \equiv A' - C, \hspace{1.0cm}
 \Delta_d^{(0)} = a_d \equiv A' + C.
\end{equation}
The first corrections in perturbation theory are
\begin{equation} \label{28}
\Delta_p^{(1)} =
\langle g_p^{(0)} |\lambda \gamma_{\bf k}^2 |g_p^{(0)} \rangle =
 {3\over 16} \lambda,  \hspace{1.0cm}
\Delta_d^{(1)} =
 \langle g_d^{(0)} |\lambda \gamma_{\bf k}^2 |g_d^{(0)} \rangle =
 {1\over 16} \lambda.
\end{equation}
At $\lambda \ge 1$ the corrections become too large
and the perturbation theory is not valid.
However, general solutions of Eq. (\ref{26}) are obvious:
\begin{eqnarray}
\label{29}
&& p\mbox{-wave}: \ g_{\bf k} \propto
  {{\sin k_x} \over {\Delta_p-\lambda \gamma_{\bf k}^2}},\\
&& d\mbox{-wave}: \ g_{\bf k} \propto
  {{\cos k_x-\cos k_y} \over {\Delta_d-\lambda \gamma_{\bf k}^2}}.\nonumber
\end{eqnarray}
Let us emphasize that the factor $(\Delta_s-\lambda \gamma_{\bf k}^2)^{-1}$
 ($s$ is the symmetry label, $s=p,d$)
 substantially enhances the wave function near the edge of the Brillouin zone
 ($\gamma_{\bf k}=0$).
Substitution of the solution (\ref{29}) into the BS equation (\ref{26}) gives
 the selfconsistency condition for $\Delta_s$.
The binding energy vanishes at the point $t=t_s$ where $a_s$ (\ref{27})
 vanishes. Near this point
\begin{equation}
\label{30}
  \Delta_s \sim - {{a_s^2}\over{\lambda}} \sim -(t-t_s)^2
\end{equation}
To be precise, we should note that for the $d$-wave there is an additional weak
 logarithmic ($\ln { {\lambda} \over{|t-t_d|} }$) dependence
 in the coefficient in this estimation.

 The numerical solutions for $\Delta_s$ as a function of $t$ are presented in
Fig. 6 ($d$-wave) and Fig. 7 ($p$-wave).  Dashed lines correspond to
$\alpha=1$ in vertex (\ref{24}) which is the physical value.  There are higher
orders in $\alpha$ which we do not take into account.  To estimate their
contribution, we present also results with $\alpha=0.5$ (solid lines). We
believe that exact solution lies somewhere between $\alpha=0.5$ and
$\alpha=1$. Of course the choice of $\alpha=0.5$ is rather arbitrary. However,
we can recall the linear expansion in $\alpha$ for the parameters $x$ and $y$
which determine the single-hole dispersion (\ref{3}): $x=-12q_1=\alpha$,
$y=-4q_1=\alpha/3$ \cite{Sus2}.
To fit the physical values (\ref{6}), we must take $\alpha \approx 0.5$.
Anyway, it is not very important since the interval between the
 curves with $\alpha=0.5$ and $\alpha=1$ in Figs. (6) and (7) is quite narrow.
At most $t$, the $d$-state has lower energy.
There is a good agreement between our results and results obtained by a
 numerical variational method \cite{Ede2} which are also presented
 in Figs. (6) and (7).
Thus to our view, the bound state disappears at $t \approx 2-3$.
This conclusion is in agreement with the results of recent works
 \cite{Bon2,Poilblanc}.

  Solving the BS equation for ${\bf p}\not= 0$, one can easily derive the
dispersion of the bound state. For example for small ${\bf p}$ at $t=J$
\begin{equation}
\label{31}
 \delta E_p({\bf p}) \approx 0.165{\bf p}^2, \hspace{1.cm}
 \delta E_d({\bf p}) \approx 0.048{\bf p}^2.
\end{equation}
For ${\bf p}\not= 0$ the solution in general case has no definite
symmetry ($d$ or $p$). The labels in above equation denote the symmetry of
corresponding solution at ${\bf p}=0$.

\section{Conclusion}

We calculated the vertex function of contact interaction between two
 holes in the $t$-$J$ model.
Contact interaction is caused by exchange of spin
 excitations with momentum $q \sim \pi$ and has no retardation.
We considered this interaction for both
 parallel and opposite directions of the holes spins.
For the most interesting case of opposite spins, there was found
 an attraction at $t/J \le 2-3$ which gives a very shallow bound state
 (with $p$ or $d$ symmetry).
Our result for binding energy agrees with that of recent works
 \cite{Ede2,Bon2,Poilblanc}, but disagrees with that of exact diagonalizations
 on small-size clusters \cite{Bon9},\cite{Has9}-\cite{Rie1}.
Following Refs. \cite{Bon9,Ede2,Pre0}, we suspect that the
 discrepancy is caused by finite-size effects in exact diagonalizations.
Recent calculation \cite{Poilblanc} of the binding energy
 for larger clusters (up to 26 sites) and the scaling with system
 size probably confirms this point of view.

The real physical value of $t/J$ in high-$T_c$ superconductors is
 $t/J \approx 3$ (see, e.g., \cite{Esk0,Fla1,BCh3}).
Therefore, the two hole short-range bound state obtained in the
 present work is probably irrelevant to high-$T_c$ superconductivity.
Moreover, we think that this bound state is by-product
 of the pure $t$-$J$ model.
Actually, the size of bound state is of the order of one lattice space
 $a \sim 7a_B$ ($a_B$ is the Bohr radius).
Therefore, the Coulomb hole-hole repulsion on neighbor sites is
 $V \sim e^2/(\epsilon a) \approx 4 $eV$/\epsilon$, where $\epsilon$
 is the dielectric constant. More realistic structure of hole wave function,
 which takes into account oxygen ions \cite{Fla1}, does not change this
 estimation.
Even with the static value $\epsilon \sim 50$, the Coulomb
 repulsion $V \sim 0.1$ eV destroys the bound state
$\Delta_d \sim -0.3J \sim -0.03$ eV. Actually the Coulomb repulsion is
even larger because in interaction we must use the dynamic value of the
dielectric constant $\epsilon(\omega)$ with $\omega \sim T_c \sim 0.01$ eV
which is definitely smaller than the static value.

Thus we have to introduce a short range hole-hole repulsion into the $t$-$J$
model. One way to do it is to consider a $t$-$J$-$V$ model \cite{Kiv0,DaR2}.
However, we do not think that the results are sensitive to the specific way of
introduction of short range repulsion.  We believe that the main problem is
long range dynamics at distances $\sim p_F^{-1}$. This problem includes the
long range antiferromagnetic order instability \cite{Sus3}.  Nevertheless the
short range repulsion is very important.

The situation is quite similar to the pion condensation in nuclear matter
\cite{Mig8}, where both the interaction of a nucleon with the Goldstone
excitation (pion) and short range nucleon-nucleon repulsion are important.
At the moment we believe that suitable description of high-$T_c$
superconductors is an effective long-range theory with spin-1/2 holes and
Goldstone gapless spin-waves.
(To avoid misunderstanding, we should note that long range instability
 destroys antiferromagnetic order and therefore Goldstone spin-waves
 definitely are not the physical excitations.
 They just provide a suitable basis set.)
The single hole properties as well as hole - spin-wave vertex \cite{Suh3}
are described by the $t$-$J$ model.  However besides that, we have to
introduce ``by hands'' the contact hole-hole repulsion.

It is interesting to notice that investigation of high-$T_c$ superconductivity
could help to better understanding of pion condensation.  Actually the
long-range instability (pion condensation) \cite{Mig8} and the BCS-type
nucleon-nucleon pairing \cite{Boh9} are usually considered as independent
phenomena.  However now, due to high-$T_c$ superconductivity, we
understand that they are closely related.

{\bf ACKNOWLEDGMENTS}

  We are grateful to V. I. Belinicher who draw our attention to contact
hole-hole interaction. We wish to acknowledge stimulating discussions with
him as well as with I. B. Khriplovich and I. V. Kolokolov.  We are
especially grateful to D. Poilblanc for supplying us with some references
and communicating to us the results of work \cite{Poilblanc} prior to
publication. One of us (A.L.C.) thanks the Soros-Akademgorodok Foundation
for financial support.

\newpage
{\bf FIGURE CAPTIONS}

FIG. 1.\hspace{0.7cm} Contact interaction of the holes with parallel spins.
Only the configurations with holes in the centers of square clusters
are shown, but each hole hops within the corresponding cluster.

FIG. 2.\hspace{0.7cm} Configuration with closed holes which minimizes the
$H_J$ and gives a very small attraction between holes with parallel spins.

FIG. 3.\hspace{0.7cm} Matrix elements for the holes with opposite spins.
Only the configurations with holes in the centers of square clusters
are shown, but each hole hops within the corresponding cluster.
Fig. a represents diagonal interaction, and Figs. b,c represent hopping of
cluster.

FIG. 4.\hspace{0.7cm} Distant configurations for interaction of opposite
spin holes in the $t$-$J_z$ model. Only the configurations with holes in the
centers of square clusters are shown, but each hole hops within the
corresponding cluster.

FIG. 5.\hspace{0.7cm} Hole-hole binding energy in the $t$-$J_z$ model as a
function of $t$. The $d$- and $p$-waves are degenerate.
Solid line is the result of the present work [Eq. (\ref{18})]. The circles
are results from numerical variational method \cite{Ede2},
and the triangles are results from exact diagonalizations \cite{Bon9}).

FIG. 6.\hspace{0.7cm} Hole-hole $d$-wave binding energy in the $t$-$J$ model.
Solid line is the result of the present work at $\alpha=0.5$. Dashed
line is the same at $\alpha=1$. The circles represent the results of
numerical variational method \cite{Ede2}.

FIG. 7.\hspace{0.7cm} Hole-hole $p$-wave binding energy in the $t$-$J$ model.
Solid line is the result of the present work at $\alpha=0.5$. Dashed
line is the same at $\alpha=1$. The circles represent the results of
numerical variational method \cite{Ede2}.

\end{document}